\documentclass[english,12pt]{article}

\usepackage{setspace}
\usepackage[margin=1in]{geometry}

\usepackage{amsmath}
\usepackage{graphicx} 
\usepackage{tabularx,multirow}
\usepackage{natbib}
\newcolumntype{Y}{>{\centering\arraybackslash}X}

\usepackage{titlesec}

\titleformat*{\section}{\large\bfseries}
\titleformat*{\subsection}{\bfseries}
\titleformat*{\subsubsection}{\bfseries}

\def\bC{\textbf{C}}

\def\bJ{\textbf{J}}

\def\bS{\textbf{S}}

\def\bW{\textbf{W}}

\def\bc{\textbf{c}}

\def\bm{\textbf{m}}

\def\bx{\textbf{x}}
\def\E{\text{E}}

\def\bbeta{\boldsymbol{\beta}}
\def\btheta{\boldsymbol{\theta}}
\def\bgamma{\boldsymbol{\gamma}}
\def\bmu{\boldsymbol{\mu}}
\def\bPsi{\boldsymbol{\Psi}}
\def\bSig\mathbf{\Sigma}

\def\bSig\mathbf{\Sigma}

\begin{document}

\thispagestyle{empty}

\begin{center}
{\bf \Large The Role of Body Mass Index at Diagnosis on Black-White Disparities in Colorectal Cancer Survival: A Density Regression Mediation Approach}
\end{center}

\vspace*{2mm}
\begin{center} \textbf{
Katrina L. Devick\textsuperscript{1,*}, Linda Valeri\textsuperscript{2}, Jarvis Chen\textsuperscript{3}, Alejandro Jara\textsuperscript{4}, \\ Marie-Ab\`ele Bind\textsuperscript{5}, and Brent A. Coull\textsuperscript{1} }
\end{center}

\begin{center}
\vspace{1mm}
{\footnotesize
\noindent  $^1$Department of Biostatistics, Harvard T.H. Chan School of Public Health, Boston, Massachusetts, USA \\
$^2$Department of Biostatistics, Columbia Mailman School of Public Health, New York, New York, USA\\
$^3$Department of Social and Behavioral Sciences, Harvard T.H. Chan School of Public Health, \\ Boston, Massachusetts, USA \\
$^4$Department of Statistics, Facultad de Matem\'aticas, Pontificia Universidad Cat\'olica de Chile, Santiago, \\Chile\\
$^5$Department of Statistics, Harvard University, Cambridge, Massachusetts, USA\\

$^*$\textit{email}: kdevick@hsph.harvard.edu
}
\end{center}

\vspace*{5mm}


\noindent{\bf Summary:}

 The study of racial/ethnic inequalities in health is important to reduce the uneven burden of disease. In the case of colorectal cancer (CRC), disparities in survival among non-Hispanic Whites and Blacks are well documented, and mechanisms leading to these disparities need to be studied formally. It has also been established that body mass index (BMI) is a risk factor for developing CRC, and recent literature shows BMI at diagnosis of CRC is associated with survival. Since BMI varies by racial/ethnic group, a question that arises is whether disparities in BMI is partially responsible for observed racial/ethnic disparities in CRC survival. This paper presents new methodology to quantify the impact of the hypothetical intervention that matches the BMI distribution in the Black population to a potentially complex distributional form observed in the White population on racial/ethnic disparities in survival. We perform a simulation that shows our proposed Bayesian density regression approach performs as well as or better than current methodology allowing for a shift in the mean of the distribution only, and that standard practice of categorizing BMI leads to large biases. When applied to motivating data from the Cancer Care Outcomes Research and Surveillance (CanCORS) Consortium, our approach suggests the proposed intervention is potentially beneficial for elderly and low income Black patients, yet harmful for young and high income Black populations. 

\vspace*{7mm}\noindent{\bf Key words}: Accelerated failure time model; Cancer health disparities; Causal inference; Dependent Dirichlet process; Nonparametric Bayesian; Stochastic Interventions.

\thispagestyle{empty}

\newpage
\clearpage
\setcounter{page}{1}

\section{Introduction}
\label{s:intro}

Differences in cancer survival among racial/ethnic groups are well documented \citep{ACS}. Considering the case of colorectal cancer (CRC), disparities in survival between non-Hispanic Blacks and non-Hispanic-Whites in the United States (U.S.) manifested in the 1980s and have widened over time \citep{DeLancey2008,siegel2014}.  Since in the U.S. CRC is the third most commonly diagnosed cancer in men and women, and the third most common cause of cancer-related death \citep{ACS,WHO}, it is important to formally study the mechanisms that are causing Black-White disparities in survival to reduce the unfair, unjust, and preventable uneven burden of cancer \citep{Krieger2005}. 

It has been well established that body mass index (BMI), as defined by weight in kilograms divided by height in meters squared, is a risk factor for developing CRC. Recent literature has reported effects of BMI pre-, at-, or post-diagnosis on CRC survival. It has been hypothesized that higher BMI is associated with worse CRC prognosis, however, numerous studies have shown that the relationship is more complicated \citep{kroenke2016,kocarnik2016}. Some studies have observed better survival in overweight or class I obese patients compared to normal weight patients. Since this paradox could result from sample selection bias, reverse causality, and/or collider bias, employing causal methodology to study the relationship between race, BMI, and CRC survival is important. 

The determinants of disparities in survival for CRC patients are multifactorial.  As differences in the distributions of BMI across racial/ethnic groups are well documented \citep{ogden2006,wang2007}, and the effect of race/ethnicity on BMI has been shown to be complex \citep{wang2007}, we propose a causal inference approach based on the theory of counterfactuals \citep{rubin1974} to quantify the percentage of disparity in survival that could be reduced had Black-White disparities in BMI been eliminated in CRC patients. This framework has not been used to estimate the impact of a shift in the entire distribution of a continuous intermediate variable along the race-survival pathway.
Using the definition of non-manipulable race as proposed by VanderWeele and Robinson \citep{vanderweele2014}, we consider a stochastic intervention on BMI to determine how BMI contributes to disparities in Black-White survival among CRC patients. The implication for understanding the relationship between race and BMI on survival is needed in order to inform clinical recommendations for patients upon diagnosis of CRC, and ultimately to reduce disparity in CRC survival.

Since the relationship between race and BMI on CRC survival is complex in that the Black-White disparity differs across the range of BMI, a simple linear model may not appropriately capture the residual disparity after a hypothetical intervention in BMI. If categories of BMI representing normal-, over-weight, class I and class II/III obese individuals are made, then one can implement methodology for categorical intermediate variables \citep{valeri2016,valeri2016statmed}. This methodology estimates the impact of a hypothetical intervention in each covariate stratum that matches the probability of being in each BMI category for Blacks to that observed in Whites. However, we wish to place no restriction on the distribution of BMI and allow for a shift in the entirety of the Black BMI distribution to match a potentially complex distributional form of BMI observed in the White population, not just allowing for a shift in mean BMI or BMI categories. In order to implement a shift in the entire distribution of BMI, we consider density regression to estimate the BMI distribution. While mediation methods based on quantile regression have been proposed as an approach to move beyond shifts in the mean \citep{bind2017,Imai2010,shen2014}, the approach we take here incorporates changes in the full distribution, not only a shift in a particular quantile.

\begin{figure}[h]
  \centerline{
   \includegraphics[width=3in]{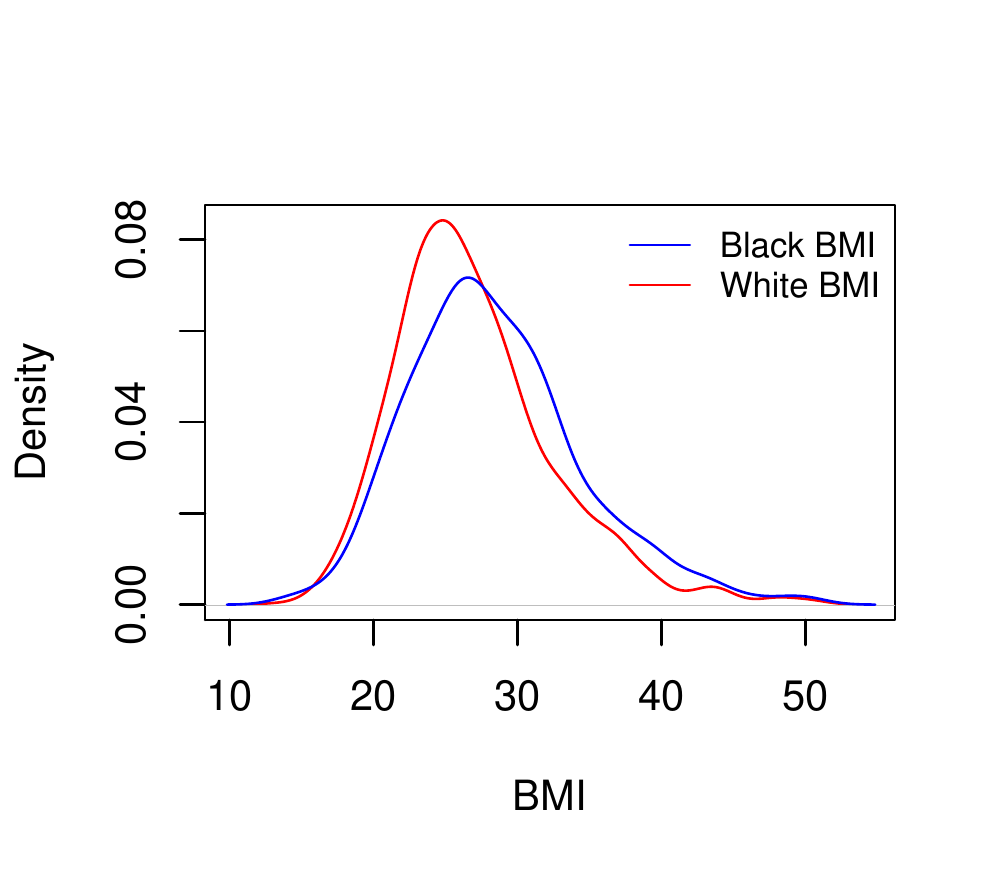}}
    \caption[Marginal BMI densities observed in CanCORS]{Marginal non-Hispanic Black and non-Hispanic White BMI distributions observed in CanCORS.}
  \label{P1:BMIden}
\end{figure}

To our knowledge, no methods exist to estimate the causal effects of a hypothetical distributional-level intervention of a continuous intermediate variable in the context of estimating racial/ethnic disparities in survival. 
As BMI is a ratio of two Gaussian random variables and therefore not itself normally distributed \citep{anderson1991,penman2006,lin2007,miljkovic2017}, we propose a Bayesian Linear Dependent Dirichlet Process (LDDP) model to nonparametrically estimate densities of BMI for each covariate stratum. We couple the LDDP model with a parametric accelerated failure time model for CRC survival to estimate the effect of this intervention. We use data from the Cancer Care Outcomes Research and Surveillance (CanCORS) Consortium to answer these questions. As shown in Figure \ref{P1:BMIden}, we observe a difference in the marginal distributions of Black and White BMI in CanCORS that is best characterized by a shift in the entire distribution, not a location shift. This dataset contains numerous individual-level variables not typically found in cancer registries and a high percentage of non-Hispanic Black CRC patients.


\section{Model}
\label{s:model}

In order to quantify the impact of a distributional shift of Black BMI to that observed in the White population, we first need to estimate the White BMI density. To nonparametrically estimate the density of White BMI by covariate pattern, and borrow strength across data with different predictor values, we consider the linear dependent Dirichlet Process (LDDP) mixture model (\ref{LDDPmodel}) \citep{deiorio2004,deiorio2009,jara2010}
\vspace*{-2mm}\begin{equation} \label{LDDPmodel}
m_i|G_{\textbf{x}_i}\overset{ind.}{\sim}\int N\left(m_i|\boldsymbol{\mu},\sigma^2\right)dG_{\textbf{x}_i}(\boldsymbol{\mu},\sigma^2),
\end{equation}
where  $m_i$ represents BMI for subject $i$, $\bx_i= \left(1, R_i, \bC_i \right)$ are the covariates in the model, and  $G_{\textbf{x}}$ is a LDDP with trajectories of the form:
$$
G_{\textbf{x}}(\cdot)= \sum_{j=1}^{\infty} \left\{V_j \prod_{l<j}(1-V_l) \right\} \delta_{\boldsymbol{\theta}_j(\textbf{x})}(\cdot),
$$
where $V_j \mid \alpha \overset{i.i.d.}{\sim}\mbox{Beta}(1,\alpha)$, $\delta_{\theta_l}(\cdot)$ is the Dirac measure at $\theta_l$, $\boldsymbol{\theta}_j(\textbf{x})=(\textbf{x}^T\boldsymbol{\beta}_j,\sigma^2_j)$, and 
$(\boldsymbol{\beta}_j, \sigma_j) \mid G_0  \overset{i.i.d.}{\sim}G_0$.  An important aspect of the LDDP mixture model is that it can be seen as a Dirichlet Process (DP) mixture of linear regression models of the form:
\begin{equation} 
m_i|G\overset{ind.}{\sim}\int N\left(m_i|\textbf{x}_i^T\boldsymbol{\beta},\sigma^2\right)dG(\bbeta,\sigma^2),
\end{equation}
\vspace*{-1.3cm}\begin{equation*}
G|\alpha, G_0 \sim DP(\alpha G_0).
\end{equation*}
This is a type of dependent Dirichlet Process (DDP), in which an uncountable set of DPs are created and dependence is introduced by modifying the stick-breaking representation of \citet{sethuraman1994}. A LDDP model has the specific restriction that the component of the atoms defining the location follows a linear regression model $\boldsymbol{\theta}_j(\bx)=(\bx^T\bbeta_j,\sigma^2_j)$.
We complete the model specification by assuming the following centering distribution $G_0 \equiv N_p(\bbeta|\bmu_b,\bS_b)\Gamma(\sigma^{-2}|\tau_1/2,\tau_2/2)$, and the following priors
\begin{equation*}
~~~~~~~~~\alpha|a_0,b_0 \sim \Gamma(a_0,b_0), \hspace*{5mm} \tau_2|\tau_{s_1},\tau_{s_2}\sim\Gamma(\tau_{s_1}/2,\tau_{s_2}/2),
\end{equation*}
\vspace*{-1.4cm}\begin{equation*}
\bmu_b|\bm_0,\bS_0 \sim N_p(\bm_0,\bS_0), \hspace*{5mm}\textnormal{and } \bS_b|\nu,\bPsi \sim IW_p(\nu,\bPsi),
\end{equation*}
where $a_0=10$, $b_0=1$, $\tau_1=6.01$, $\tau_{s_1}=6.01$, $\tau_{s_2}=2.01$, $\nu=9$, $\bm_0=(\bW^T\bW)^{-1}\bW^Ty$, $\bS_0=1,000(\bW^T\bW)^{-1}$, $\bPsi^{-1}=\bS_0$, and ${\bf W} = \left[ \bJ_n | {\bf x} \right] .$

To obtain a simple functional form of the residual disparity estimator when of the outcome is common, we model the survival time outcome via a Bayesian Weibull accelerated failure time (AFT) model. First, we estimate the Black-White disparity prior to any intervention using model  (\ref{disp}):
\begin{equation} \label{disp}
	\log(T) = \gamma_0 + \gamma_1R +  \bgamma_{\bC}^T\bC + \nu\epsilon,
\end{equation}

\noindent where $R$ is an indicator for non-Hispanic Black, $\bC$ is a matrix of confounders, $\nu$ is the scale parameter, and $\epsilon$ follows an extreme value distribution. We consider the definition by \citet{vanderweele2014} for the disparity in survival between non-Hispanic Blacks and non-Hispanic Whites prior to any intervention:
\begin{equation} 
Disparity =  \frac{\E[T|R=1, \textbf{c}]}{\E[T|R=0, \textbf{c}]}=exp(\gamma_1),
\end{equation}
which is the ratio of mean survival time in the Black population compared to the mean survival time in the White population.

 Due to evidence that BMI has a nonlinear effect on CRC survival \citep{kroenke2016,kocarnik2016}, we allow for a $q-$degree polynomial effect of BMI when modeling CRC survival. The relationship between race, BMI and CRC survival is modeled by:
\begin{equation} \label{P1:outcomeKpoly}
	log(T) = \theta_0 + \theta_1R + \sum_{j=1}^q \theta_{2j}M^j +  \sum_{j=1}^q \theta_{3j}RM^j + \btheta_{4}^T\bC +  \btheta_{5}^TR\bC + \sum_{j=1}^q \btheta_{6j}^T M^j\bC + \nu\epsilon,
\end{equation}

\noindent where $M$ represents centered continuous BMI. We show the residual disparity (RD) in survival, as defined by \citet{vanderweele2014}, between Black and White individuals after intervening to match the distribution of BMI in the Blacks to that of the Whites, for each covariate pattern is: 
\begin{equation}\label{RD}
\frac{\E\left[T_{H_{\textbf{c}}(0)} | R=1, \textbf{c} \right]}{\E\left[ T |  R=0, \textbf{c}\right]} = 
e^{\theta_1+ \btheta_{5}^T\bc }
\frac{\int_m e^{  \sum_{j=1}^q \left(\theta_{2j}+\theta_{3j}\right)m^j  + \sum_{j=1}^q \btheta_{6j}^T m^j\bc}  dP_M(m|R=0,\bc) 
}
{ \int_m e^{  \sum_{j=1}^q \theta_{2j}m^j + \sum_{j=1}^q \btheta_{6j}^T m^j\bc }  dP_M(m|R=0,\bc)},
\end{equation}

\noindent where $H_{\textbf{c}}(0)$ represents a random draw from the White BMI distribution for the corresponding covariate stratum. The RD formula given in (\ref{RD}) only requires estimation of the BMI distribution to which the intervention normalizes this intermediate variable, which we term the target distribution. Thus, only estimation of the White BMI density, our target distribution, for each covariate stratum is necessary. The RD measure assumes 1) no unmeasured mediator-survival confounding and 2) both the mediator and outcome models are correctly specified. 
See Appendix \ref{AppendixA} for the derivation of the RD formula given in (\ref{RD}).

We use the \texttt{DPpackage R} package \citep{jara2011} to fit the LDDP model for BMI. To obtain posterior samples of the conditional density and cumulative density functions (CDF) for each Markov chain Monte Carlo (MCMC) iteration of (\ref{LDDPmodel}), we add an additional Fortran function to the LDDPdensity.f file in \texttt{DPpackage}. The appended LDDPdensity.f file and \texttt{R} code implementing our density regression approach are available on GitHub \citep{devick2018}.
To estimate the integrals in (\ref{RD}), we use inverse cumulative density sampling \citep{morgan1984} to draw 1,000 random samples from these conditional BMI density curves. Using model (\ref{LDDPmodel}), we estimate a density and CDF for a grid of 200 BMI values for all White covariate strata. We linearly interpolate random draws of BMI from the grid.  To implement the shift in distribution of the Black BMI to match the White distribution of BMI, we use the White density and CDF for each covariate pattern to estimate the integral in the numerator. For each of the random samples in a particular covariate stratum, we calculate the necessary integrand and then take the mean of these integrand samples as the estimate of the integral. We do this for the densities and CDFs at each iteration of the MCMC to obtain posterior samples of the residual disparity in each covariate stratum. From these posterior samples, we calculate 95\% credible intervals. We acquire posterior samples of the marginal residual disparity by weighting the stratum specific posterior samples of the residual disparity by the empirically estimated probability of being in a given covariate stratum. Lastly, we consider the percent reduction in disparity as: 
\begin{equation}
Disparity \hspace*{.03in}Reduction(\textbf{c}) = \frac{disparity - residual \hspace*{.03in} disparity(\textbf{c})}{disparity}.
\end{equation}


\section{Simulation Study}
\label{s:sim}

\subsection{Simulation Setup} 
We used simulation to compare the operating characteristics of the residual disparity estimator from our density approach to those obtained by the traditional/difference method \citep{baron1986}, the method that assumes a linear regression model for BMI, and the method that categorizes BMI into normal-, over-weight, class I and class II/III obesity and uses a baseline category logit (BCL) regression to model categorical BMI. For the method assuming a linear regression for BMI, we performed nonparametric inference by simulating counterfactuals, as outlined by \citet{Imai2010}. Model specification and the residual disparity formula for the baseline category logit model can be found in Appendix \ref{AppendixB}.

We compared the operating characteristics of these estimators under four scenarios for race-specific BMI distributions.  First, we assumed that the true distribution of BMI was normal and there was only a location shift between the BMI distributions for Blacks and Whites. Then, we considered three additional scenarios where the difference between the Black and White BMI distributions cannot be described as a location-scale shift. In these later scenarios, the Black BMI distribution is normally distributed, but the target distribution, the White BMI distribution, is a generated by a mixture of normals.

\begin{figure}[h]
   \includegraphics[width=5.2in]{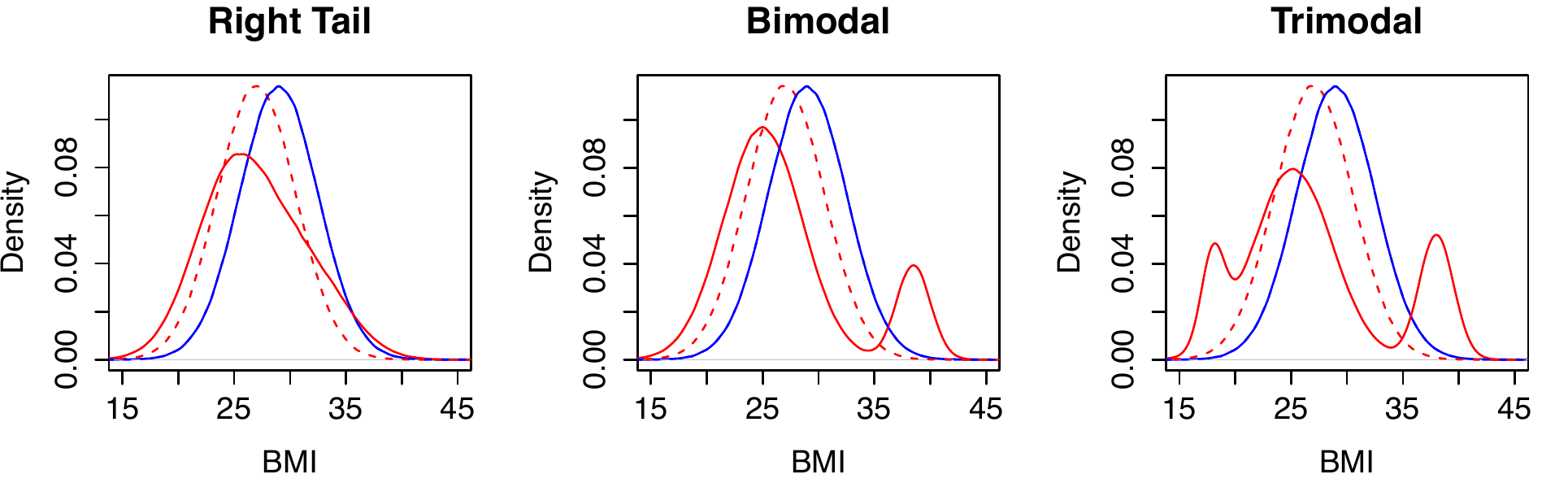} \includegraphics[width=1.25in]{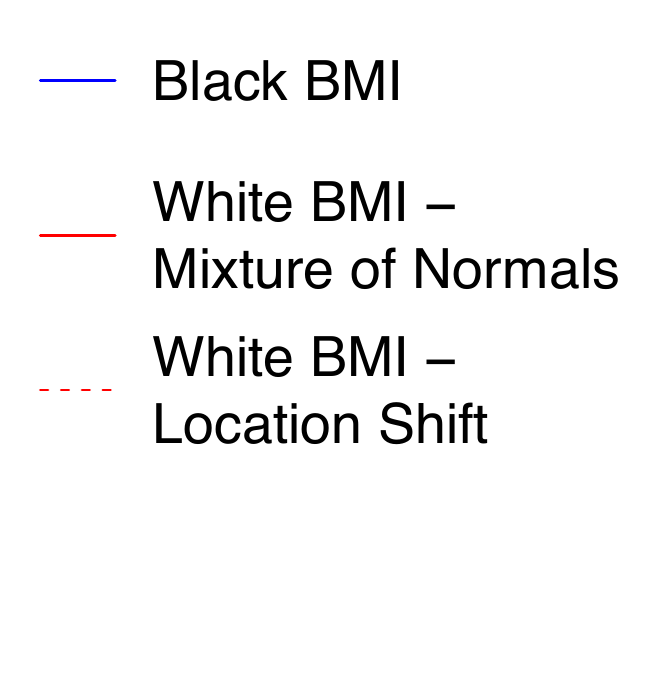}
   \caption{True distributions of BMI under three cases of our simulations. The location shift only case is represented by dashed lines. }
  \label{sim}
\end{figure}

We simulated $M$ and $T$ for $K=1,500$ subjects, letting 750 be Black ($R=1$) and 750 be White ($R=0$). We randomly sampled 750 values from the true underlying BMI distribution for the corresponding race. For the location shift only, we generated BMI values for White subjects from a Normal($27,3.5^2$). The true underlying BMI distributions for the three scenarios when the White BMI distribution was a mixture of normals are summarized in Figure \ref{sim}. For each scenario, we fixed the mean of the White BMI distribution at 27. 

Since the formula for the residual disparity estimators given in (\ref{RD}) only requires
estimation of the target distribution, we generated Black BMI values from a Normal($29,3.5^2$) in all four cases. We then simulated survival times ($T_s$) for each subject from:
\begin{equation}\label{survtimes}
log(T) = \theta_0 + \theta_1 R + \theta_2M + \theta_3RM+\nu\epsilon, 
\end{equation}
\noindent where $\nu=0.82$, $\epsilon$ is a random sample from an extreme value distribution, and we set to $\btheta=(7.56, -0.88, 0.029, 0)$ or $\btheta=(7.56, -0.88, 0.029, 0.022)$ under no interaction and interaction respectively. We generated random censoring times ($T_c$) from a Weibull distribution, such that approximately $65\%$ of the observed times, $T =$min$(T_s,T_c)$, were censored. We truncated the observed times if $T > 1826.25$ days. We simulated 500 datasets for each of the four scenarios of true underlying BMI distributions. For each dataset, we estimated the disparity and residual disparity (RD) for each of the three mediator models and under the difference method. We compared the performance of these estimators between the different scenarios of underlying distributions of BMI via root mean squared error (rMSE), bias, and standard deviation.

In order to assess how the magnitude of the deviations from standard assumptions would affect the relative performance of the methods under consideration, we repeated our simulation when the effect sizes were doubled, $\btheta=(7.56, -1.76, 0.058, 0)$ or $\btheta=(7.56, -1.76, 0.058,$ $0.044)$, under no interaction and interaction respectively.

\subsection{Simulation Results}

Table \ref{P1:simresults} shows the rMSE, bias, and standard deviation (SD) for the residual disparity estimators from the LDDP, linear regression, baseline category logit models and the difference method. The rMSE, bias, and SD values exhibit similar patterns when the $\btheta$ values are motivated from CanCORS and when they are doubled. When a RM interaction exists ($\theta_3 \ne 0$), our density regression approach performs comparably to or better than the linear regression approach. The bias is smallest when a LDDP mediator model is used, and the variance is only slightly larger than in the linear case. As the effect size increases and the underlying mediator distribution becomes less normal, larger differences among the various residual disparity estimators emerge. Notably, when the mediator is categorized as is popular in BMI research, and a baseline category logit (BCL) model is used, we observe large bias and variance. The baseline category logit and the difference method estimators perform equally badly, being about 5\% worse than our proposed density regression estimator in terms of bias and rMSE. As we expected, the most efficient method is the difference method, but this leads to large bias when interactions and nonlinearities are present. When no interaction is present ($\theta_3=0$), the four methods perform similarly. This suggest that even when more complicated methods are not needed, our proposed density regression approach performs as well as simpler methods in terms of rMSE. \\


\section{Data analysis}
\label{P1:cancors}

\subsection{Study population}

We obtained data from the National Cancer Institute's CanCORS Consortium, which contains detailed information from patients and physicians \citep{ayanian2004}. The consortium collected data on CRC cases in multiple regions and health care delivery systems across the U.S. The study population consisted of non-Hispanic White and non-Hispanic Black patients enrolled between 2003 and 2005. To be eligible for enrollment as a CRC case, the study required patients to be at least 21 years old and with newly diagnosed invasive adenocarcinoma of the colon or rectum. The CanCORS Consortium oversampled minority groups at several of the sites. Of the ten CanCORS centers that enrolled patients with CRC, 8 centers collected survival data, including: Henry Ford Health System (HFHS), Kaiser Permanente Hawaii (KPHI), Kaiser Permanente Northwest (KPNW), Northern California Cancer Center (NCCC), State of Alabama (UAB), Los Angeles County (UCLA), North Carolina (UNC), and Veterans Health Administration (VA). We excluded patients from Kaiser Permanente Hawaii from our analysis due to a small number of non-Hispanic Blacks. 

Patients self-reported height and weight at diagnosis of CRC during the CanCORS survey (approximately 4-6 months after diagnosis). We used these measurements to calculate BMI, as defined by weight in kilograms divided by height in meters squared. CanCORS collected stage at diagnosis information via medical record abstraction or from cancer registries, categorized as stage I-IV according to the American Joint Committee on Cancer (AJCC) staging criteria \citep{ajcc}. For the purpose of the analysis, we excluded patients for whom stage was unknown (approximately 7\% of the sample). 

\subsection{Models}

For our CanCORS data analysis, we used survival time in months since diagnosis, and found after restricting to 60-month survival, 62\% of the Black and 70\% of the White populations were censored. We considered categorized age at diagnosis (age $< $50, 50-65, and $>$65 years), sex (female versus male), income level ($<$\$40$k$, \$40-60$k$, \$60-80$k$, $>$\$80$k$), and stage at diagnosis (I-IV) to be potential confounders of the BMI-survival relationship. We compared the distribution of baseline characteristics for non-Hispanic Blacks to non-Hispanic Whites via $\chi^2$ statistics. 

In CanCORS, we estimated the Black-White disparity in CRC survival prior to intervention on BMI, and the residual disparity after hypothetical intervention on the distribution of Black BMI to match the observed distribution of White BMI using our density approach, linear regression for BMI, BCL regression for categorical BMI, and the traditional/difference method. Based on exploratory analyses, we allowed for a quadratic effect of BMI on CRC survival in the LDDP, linear regression, and BCL mediator models. We calculated 95\% confidence intervals for the disparity, and conditional and marginal residual disparity measures for the linear and BCL models via the bootstrap. We investigated heterogeneity in the effect of the hypothetical intervention on Black BMI in CanCORS across covariate strata. 

Motivated by \citet{valeri2016}, we conducted backwards model selection allowing for race by age, race by sex, BMI by age, and BMI by sex interactions. We performed analyses using the best fitting models for BMI and CRC survival in CanCORS, and compared these results to those that did not include any of these interactions.

\subsection{Results}

After we apply exclusion criteria, our sample from CanCORS consisted of 1,585 CRC patients (1,253 non-Hispanic White and 332 non-Hispanic Black individuals). Black and White individuals differ significantly on many baseline characteristics. There are more elderly Black individuals than White, Black patients tend to be poorer than White patients, and Black CRC patients are less likely to survive 5 years (log rank test $p$=0.005).  

\begin{figure}
  \centering
   \includegraphics[width=5in]{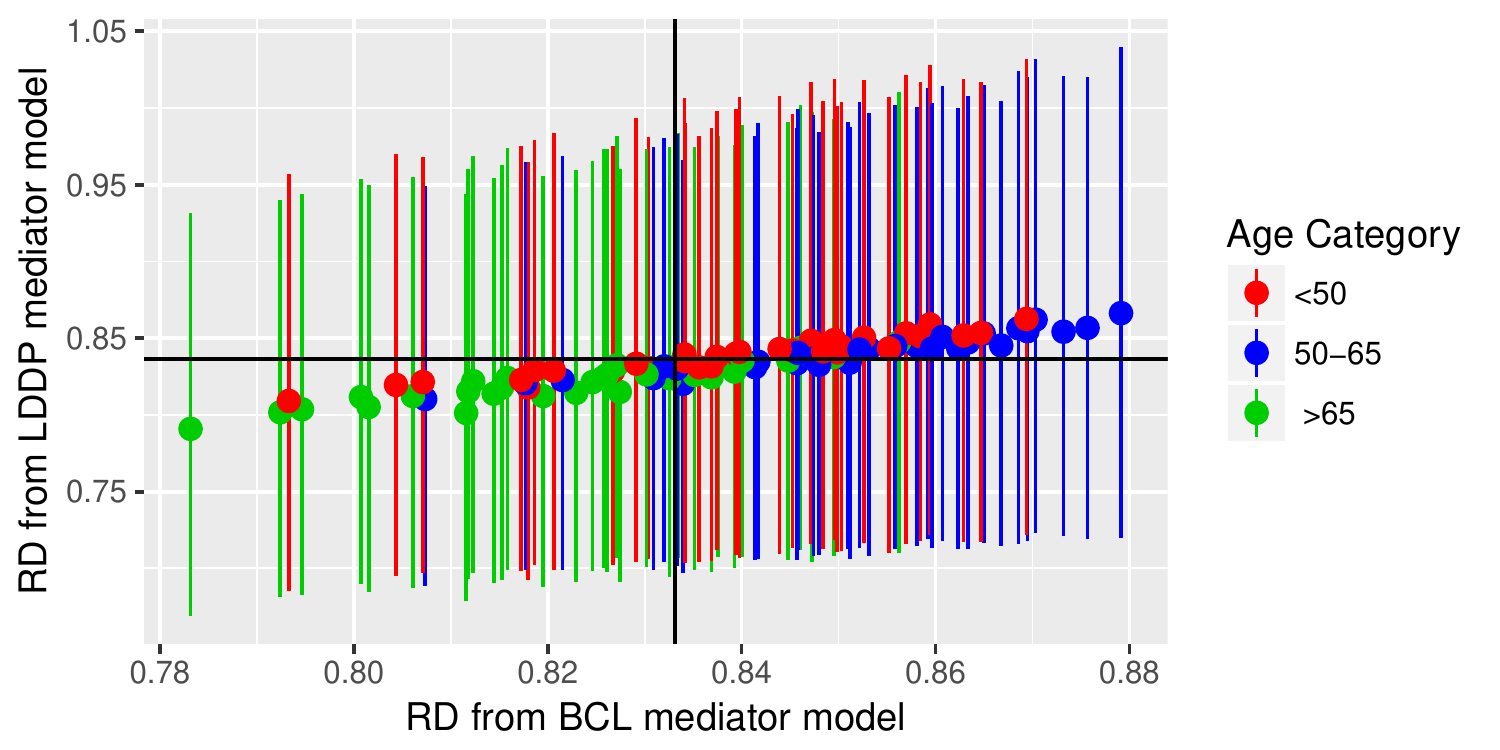}\\
   
    \vspace*{6mm}\includegraphics[width=5in]{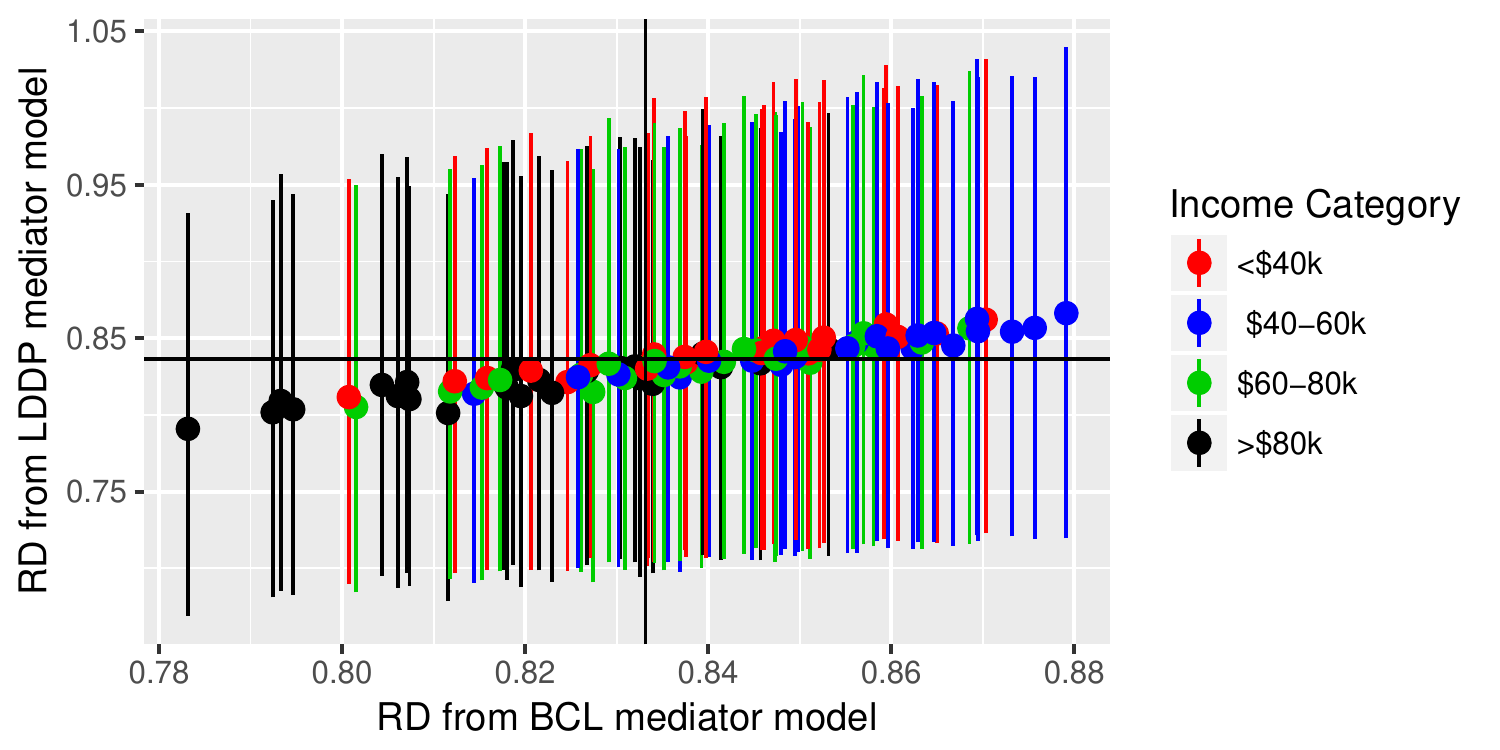}
     \caption[Residual disparity estimates from the BCL regression approach compared to our density regression approach.]{Residual disparity estimates from the BCL regression approach compared to our density regression approach, without considering potential race$\times$covariate and BMI$\times$covariate interactions. Each point represents the residual disparity for a particular covariate pattern. Estimated disparity in survival prior to intervention are plotted as vertical and horizontal lines. 95\% credible intervals from our density regression approach are depicted in the figure.}
  \label{P1:RDbmisq}
\end{figure}

Before any hypothetical intervention on BMI at diagnosis, we observe a significant Black-White disparity in survival, adjusting for covariates ($p=0.023$). We explore if differences in the impact of the intervention exist across covariate strata by examining conditional RD estimates. Figure \ref{P1:RDbmisq} compares conclusions drawn from the BCL approach to our density regression approach for each covariate pattern, when the AFT model does not include race-covariate interactions. The disparity in survival estimates prior to intervention are plotted as vertical and horizontal lines. Covariate patterns for which both methods estimate the effect of the intervention of BMI to be harmful are in the bottom left quadrant, and points for which both methods estimate the effect to be beneficial are in the upper right quadrant. Points in the upper left and lower right quadrants are covariate patterns for which the two methods lead to different conclusions about the intervention on BMI. 95\% credible intervals estimated from the density regression approach are included on the plots. We see heterogeneity in the impact of intervention on BMI exists across subpopulations in CanCORS, as there are covariate strata for which the intervention is beneficial and some for which the interventional is harmful, although no differences are significant. To examine trends in the RD estimates across covariates, we colored one plot by age categories and another by income categories. 

\begin{figure}
  \centering
   \includegraphics[width=5in]{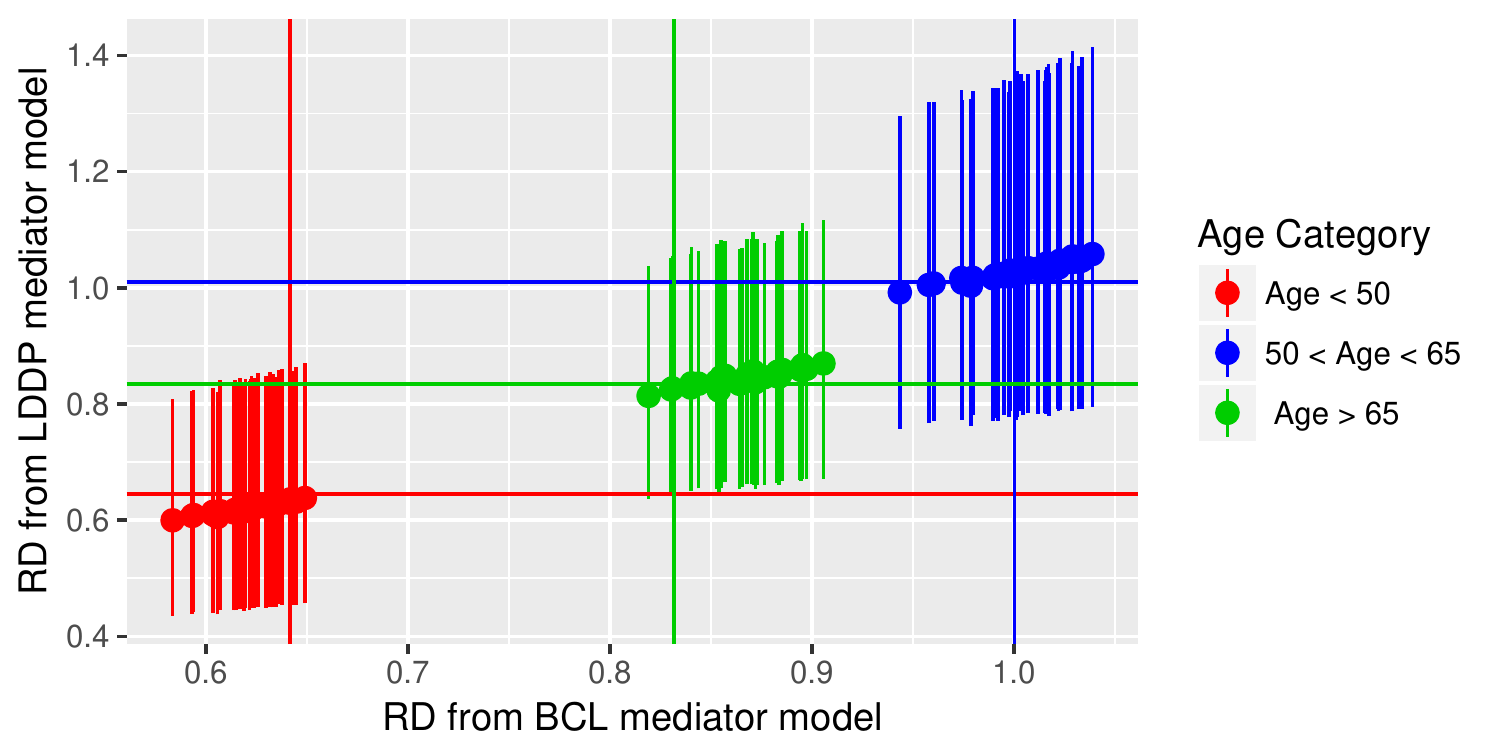}\\
     \hspace*{1mm} \includegraphics[width=5.1in]{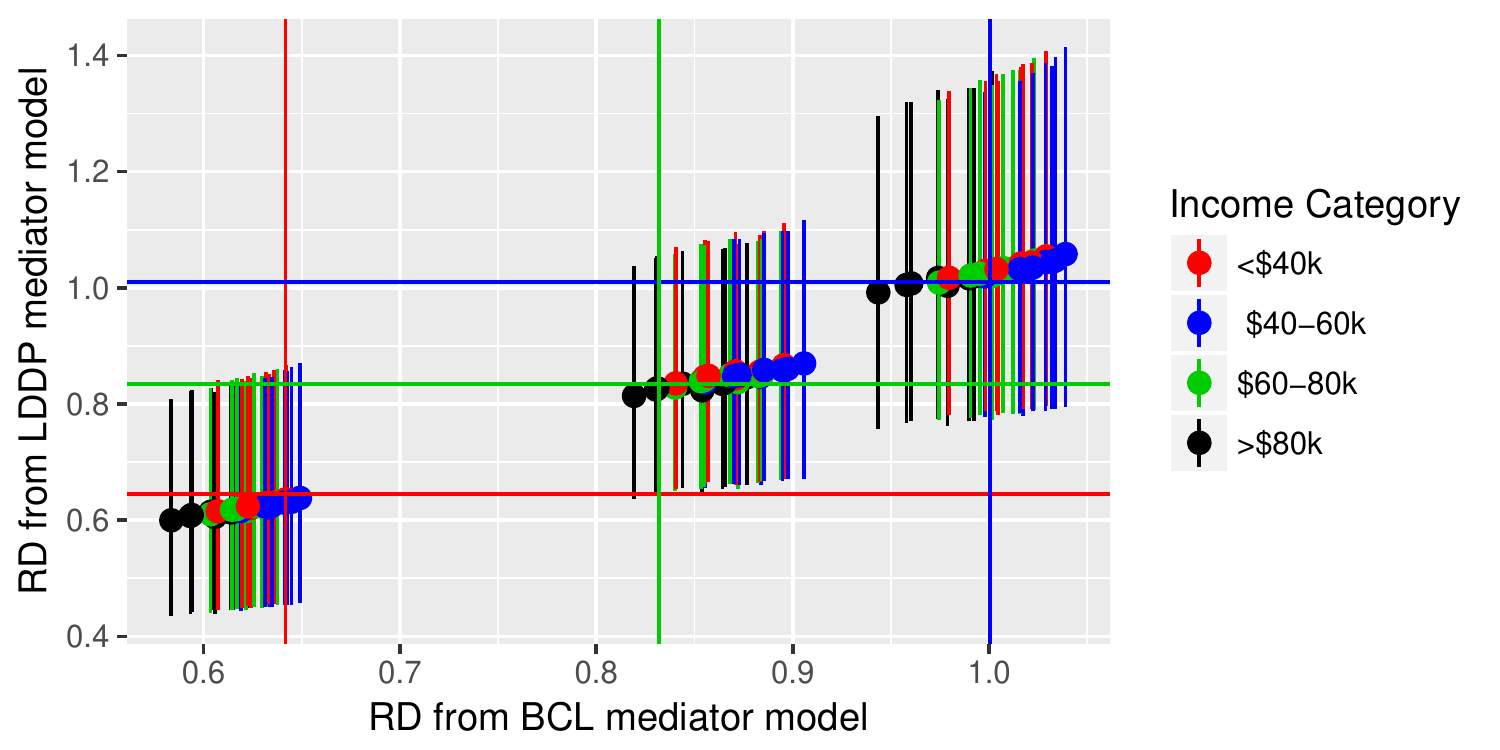}\\
    \caption[Residual disparity estimates from the BCL regression approach compared to our density regression approach for the best fitting model.]{Residual disparity estimates from the BCL regression approach compared to our density regression approach for the best fitting mediator and outcome models. The best fitting model for BMI included a race-sex interaction and the best fitting survival model included race-age interactions. Each point represents the residual disparity for a particular covariate pattern. Estimated disparity in survival for each age group prior to intervention are plotted as vertical and horizontal lines.}
      \label{P1:RDbest}
\end{figure}

In the CanCORS population, we find race-age interaction terms in the AFT model for the joint effect of race and BMI on CRC survival are statistically significant (LRT $p = 0.050$) and a race-sex interaction term in the linear regression model for BMI is significant (LRT $p<0.001$). Race-age interaction terms in the AFT model for the total effect of race on CRC survival are marginally significant (LRT $p=0.098$). This suggests that age-specific disparities should be used when determining the impact of an intervention on BMI. Results from each of the methods after re-running our analyses with these significant interaction terms are summarized in Table \ref{P1:tableBEST} and Figure \ref{P1:RDbest}. The marginal residual disparity accounting for race-age and race-sex interactions in Table \ref{P1:tableBEST} are almost identical to the estimated residual disparity not accounting for these interactions. However, due to the large effect size of the race-age interactions in the AFT model, we observe gaps in the stratum specific residual disparity and disparity estimates. After allowing for race-covariate interactions, the trend is such that the intervention is harmful for younger patients, yet beneficial for the elderly population. This trend is different than what we see in Figure \ref{P1:RDbmisq}, suggesting that the conditional results can be sensitive to race-covariate interactions in the survival outcome and disparity models. The impact of the intervention according to income categories remains robust to model misspecification, as the ordering of the RD estimates by income categories, across age groups, remains the same. When comparing the RD estimates in each age category to the corresponding age category specific disparity, we see the impact of the intervention to match the distribution of Black BMI to the distribution of White BMI is not significant for any covariate strata.

%
%


\section{Discussion}
\label{P1:discuss}

To our knowledge, this is the first study to implement a counterfactual causal inference approach to estimate the impact of a shift in the distribution of a continuous intermediate variable of arbitrary form along the race-survival pathway in survival disparities. Our methodology relies on the assumptions that there are no unmeasured BMI-survival confounders and that our models are correctly specified. 

The main conclusions of our paper are two-fold. First, we show it is not only inefficient to categorize a mediator variable, but categorization results in more bias and variance than if the continuous analog was used, even if the distribution of the mediator is not normal. This is particularly important when the intermediate variable is BMI, since categorizing BMI into well-established categories of normal-, over-weight, and obese is standard in the biomedical literature. One way to avoid the potential for large biases in such settings is to use the nonparametric regression methods proposed here. Second, when race-covariate interactions exist and are not accounted for, incorrect conclusions can be drawn about the impact of intervention on the intermediate variable in reducing race-survival disparities. 

Our approach had several advantages. The density mediator model in our approach is robust to model misspecification. This is a consequence of the BMI densities being estimated nonparametrically for each covariate pattern. Also, even with the loss of efficiency from a nonparametric mediator model, our simulations show the reduction in bias can outweigh any increase in efficiency, resulting in decreases in rMSE. Our density regression approach is the only method that allows for a arbitrary shift in the distribution of a continuous intermediate variable, as opposed to a simple location shift. In situations in which there are unmeasured variables, which when marginalized over lead to complex, multi-model distributional forms for the mediator, or when the sample size is insufficient to stratify on all observed variables that would reduce the potential for such complex distributions, our proposed methodology allows the distribution of the mediator to be flexible. 

There are limitations to our analysis. Since the data are sparse, race by income interactions were not considered and additional covariates were not added to our models. Thus, estimates could be biased due to residual confounding. Although, our methods are robust to the exclusion of covariates in which omission results in a non-normal distribution of the mediator, since conditional densities are estimated nonparametrically. In the case of our data application to CanCORS, we only have a slightly right tailed distribution of BMI, which does not deviate enough from a normal distribution to see significant differences between RD estimates from the various approaches in our results. For some covariate strata, the White BMI distribution is less advantageous to the Black BMI distribution; therefore, the universal intervention of matching the Black mediator distribution to the White mediator distribution may not be the best intervention when BMI is considered as the intermediate variable. Even though CanCORS contained individual level income, due to low numbers of Black individuals in higher income categories, we needed to the use crude income categories. Ideally, we would have data on BMI one year after diagnosis of CRC in addition to BMI at diagnosis of CRC, so individual trajectories of weight loss/gain could be considered. The major limitation in our analysis is the small number of non-Hispanic Blacks included in CanCORS after applying exclusion criteria (n=332). 

Further study is needed to more thoroughly understand the relationship between race-survival and moderating factors. Methods that consider weighting approaches will allow for more covariates and interactions to be included in the analysis. Extensions to our approach to allow for dependence between the mediator and outcome models will be considered in future research. In this paper, we consider the universal shift in the distribution of a continuous intermediate variable in each covariate stratum that matches the distribution of the mediator in the Black population to that of the White population. Future research could implement different interventions across covariate strata and/or consider interventions on both the Black and White intermediate variable distributions.

\appendix
\numberwithin{equation}{section}


\newpage
\noindent{\Large\bf Appendix}
\section{General residual disparity formula derivation}
\label{AppendixA}

\noindent  The general formula to estimate the residual disparity in covariate stratum $\bc$ for any mediator model, assuming (\ref{P1:outcomeA}) holds, is (\ref{P1:RDgenA}). This formula allows for all possible race-covariate interactions, mediator-covariate interactions, and a possible $q$-degree polynomial effect of the mediator (M) on survival.
\begin{equation} \label{P1:outcomeA}
	log(T) = \theta_0 + \theta_1R + \sum_{j=1}^q \theta_{2j}M^j +  \sum_{j=1}^q \theta_{3j}RM^j + \btheta_{4}^T\bC +  \btheta_{5}^TR\bC + \sum_{j=1}^q \btheta_{6j}^T M^j\bC + \nu\epsilon
\end{equation}

\noindent $\E_T(T_{H_{\textbf{c}}(0)} |  R=1, \bC=\bc)$ represents the expected survival time for a Black individual with covariate levels $\bc$ if their BMI at diagnosis of colorectal cancer had been set to a random draw from that of the white BMI distribution with covariate levels $\bc$. \\

\noindent $\E_T(T_{H_{\textbf{c}}(0)} |  R=1, \bC=\bc)$ 
	\vspace*{-5mm}\begin{eqnarray*}
		 &=& \int_m \E_T(T_{m} |  R=1, H_{\textbf{c}}(0)=m, \bC=\bc) dP_{H_{\textbf{c}}(0)}(m|R=1,\bc)\\
&=& \int_m \E_T(T_{m} |  R=1, \bC=\bc) dP_{H_{\textbf{c}}(0)}(m|R=1,\bc)\\
		&=& \int_m \E_T(T |  R=1, M=m, \bC=\bc) dP_M(m|R=0,\bc)\\
	&=& \int_m \E_T\left[ e^{  \theta_0 + \theta_1 + \sum_{j=1}^q \left(\theta_{2j}+\theta_{3j}\right)m^j + \left(\btheta_{4}^T+\btheta_{5}^T\right)\bc + \sum_{j=1}^q \btheta_{6j}^T m^j\bc + \nu\epsilon} \right] dP_M(m|R=0,\bc)\\
	&=& \int_m e^{  \theta_0 + \theta_1 + \sum_{j=1}^q \left(\theta_{2j}+\theta_{3j}\right)m^j + \left(\btheta_{4}^T+\btheta_{5}^T\right)\bc+ \sum_{j=1}^q \btheta_{6j}^T m^j\bc  }  \E_T\left[ \nu\epsilon  \right] dP_M(m|R=0,\bc)\\
	&=& e^{ \theta_0 + \theta_1 + \left(\btheta_{4}^T+\btheta_{5}^T\right)\bc  }   \int_m e^{  \sum_{j=1}^q \left(\theta_{2j}+\theta_{3j}\right)m^j  + \sum_{j=1}^q \btheta_{6j}^T m^j\bc}  dP_M(m|R=0,\bc) \\
\end{eqnarray*}

	\noindent$\E_T(T_m |  R=0, \bC=\bc)$
	\vspace*{-5mm}\begin{eqnarray*}
		&=& \int_m \E_T(T_m |  R=0, M=m, \bC=\bc) dP_M(m|R=0,\bc)\\
&=& \int_m \E_T(T_m |  R=0, \bC=\bc) dP_M(m|R=0,\bc)\\
		&=& \int_m \E_T(T |  R=0, M=m, \bC=\bc) dP_M(m|R=0,\bc)\\
		&=& \int_m \E_T\left[ e^{  \theta_0 + \sum_{j=1}^q \theta_{2j}m^j + \btheta_{4}^T\bc  + \sum_{j=1}^q \btheta_{6j}^T m^j\bc + \nu\epsilon} \right] dP_M(m|R=0,\bc)\\
	&=& \int_m e^{  \theta_0 + \sum_{j=1}^q \theta_{2j}m^j + \btheta_{4}^T\bc  + \sum_{j=1}^q \btheta_{6j}^T m^j\bc }  \E_T\left[ \nu\epsilon  \right] dP_M(m|R=0,\bc)\\
	&=& e^{ \theta_0  + \btheta_{4}^T\bc  }  \E_T\left[e^{  \nu\epsilon }\right]   \int_m e^{  \sum_{j=1}^q \theta_{2j}m^j + \sum_{j=1}^q \btheta_{6j}^T m^j\bc }  dP_M(m|R=0,\bc) 
	\end{eqnarray*} 

 \vspace*{-1cm}\begin{equation}\label{P1:RDgenA}
RD(\bc) = \frac{\E\left[T_{H_{\textbf{c}}(0)} | R=1, \textbf{c} \right]}{\E\left[ T |  R=0, \textbf{c}\right]} = 
e^{\theta_1+ \btheta_{5}^T\bc }
\frac{\int_m e^{  \sum_{j=1}^q \left(\theta_{2j}+\theta_{3j}\right)m^j  + \sum_{j=1}^q \btheta_{6j}^T m^j\bc}  dP_M(m|R=0,\bc) 
}
{ \int_m e^{  \sum_{j=1}^q \theta_{2j}m^j + \sum_{j=1}^q \btheta_{6j}^T m^j\bc }  dP_M(m|R=0,\bc)}
\end{equation}


\section{Residual disparity formula for BCL mediator model}
\label{AppendixB}

\noindent Now, let us categorize our mediator ($M$) into $K+1$ categories. Consider the following baseline category logit (BCL) mediator model (\ref{P1:BCL}) and AFT Weibull outcome model (\ref{P1:AFTbcl}):

\vspace*{-5mm}\begin{equation}\label{P1:BCL}
\text{log}\left\{ P(M=k | R, \bC)/P(M=0 | R, \bC) \right\} = \beta_{0k} + \beta_{1k}R + \bbeta_{2k}^T\bC  + \btheta_{3k}^TR\bC
\end{equation}
\vspace*{-8mm}\begin{equation}\label{P1:AFTbcl}
 \text{log}(T) = \theta_0 + \theta_1R + \sum_{k=1}^K \theta_{2k}M_k +  \sum_{k=1}^K \theta_{3k}RM_k + \btheta_{4}^T\bC +  \btheta_{5}^TR\bC + \sum_{k=1}^K \btheta_{6k}^T M_k\bC + \nu\epsilon,
 \end{equation}
 
\vspace*{2mm}\noindent where $M_k$ is an indicator that the mediator value is in category $k$. Under these models, the residual disparity estimator for a particular covariate pattern $\bc$ is:

\vspace*{-7mm}\begin{equation*}\label{P1:RDbcl}
\frac{\E\left[T_{H_{\textbf{c}}(0)} | R=1, \textbf{c} \right]}{\E\left[ T |  R=0, \textbf{c}\right]} = 
e^{\theta_1+ \btheta_{5}^T\bc }
 \left\{ \dfrac{1 + e^{\beta_{01} + \bbeta_{21}^T\bc + \theta_{21} + \theta_{31} + \btheta_{61}^T\bc} + \hdots + e^{\beta_{0K} + \bbeta_{2K}^T\bc + \theta_{2K} + \theta_{3K} + \btheta_{6K}^T\bc} } 
 {1 + e^{\beta_{01} + \bbeta_{21}^T\bc + \theta_{21} + \btheta_{61}^T\bc } +  \hdots + e^{\beta_{0K} + \bbeta_{2K}^T\bc + \theta_{2K} + \btheta_{6K}^T\bc } }\right\}.
\end{equation*}

\newpage

\vspace*{2cm}


\vspace*{2cm}

\begin{table}[p!]
\small
\center
\caption[Simulations results when there is an interaction is present ($\theta_3 \ne 0$), when values are motivated from CanCORS and effect sizes are doubled. ]{Simulations results when there is an interaction is present ($\theta_3 \ne 0$), when values are motivated from CanCORS, and effect sizes are doubled. Root mean squared error (rMSE), bias, and standard deviation (SD) of the residual disparity estimators under LDDP, linear, baseline category logit (BCL) mediator models and the difference method for 3 different cases of true underlying White BMI distribution is right tailed (RT), binomial (Bi), or trinomial (Tri). 
}
\begin{tabularx}{\textwidth}{|c|c *{6}{|Y}|}
   \cline{2-8}
\multicolumn{1}{c|}{}&\multirow{2}{*}{\textbf{Mediator Model}} & \multicolumn{3}{c|}{$\btheta=(7.56,-0.88,0.029,0.022)$} &   \multicolumn{3}{c|}{$\btheta=(7.56,-1.76,0.058,0.044)$} \\
 \cline{3-8}
\multicolumn{1}{c|}{}& & RT  & Bi  & Tri  & RT  & Bi  & Tri  \\ 
  \cline{2-8}\hline
\multirow{4}{*}{\textbf{rMSE}} 
&Density & 0.076 & 0.079 & 0.083 & 0.112 & 0.149 & 0.158 \\ 
& Linear & 0.077 & 0.083 & 0.088 & 0.111 & 0.155 & 0.168\\ 
 & BCL & 0.078 & 0.094 & 0.095 &  0.122 & 0.194 & 0.209\\ 
 & Traditional & 0.079 & 0.088 & 0.094 &  0.141 & 0.203 & 0.229\\ 
   \hline
  \multirow{4}{*}{\textbf{Bias}} 
  &  Density & -0.032 & -0.034 & -0.034 &  -0.076 & -0.115 & -0.124\\ 
  & Linear & -0.038& -0.047 & -0.051 &  -0.075 & -0.127 & -0.141\\ 
 & BCL & -0.026 & -0.059 & -0.056&  -0.069& -0.165 & -0.175 \\ 
  &Traditional & -0.041 & -0.053 & -0.061 &  -0.120 & -0.187 & -0.214\\
   \hline
   
  \multirow{4}{*}{\textbf{SD}} 
  & Density & 0.068 & 0.072 & 0.076& 0.082 & 0.095 & 0.098\\ 
 & Linear & 0.067 & 0.069 & 0.072&  0.082 & 0.089 & 0.091\\ 
 & BCL & 0.074 & 0.074 & 0.077& 0.101 & 0.102 & 0.114 \\ 
 & Traditional & 0.067 & 0.070 & 0.072 &0.075 & 0.079 & 0.081 \\ 
 \hline
\end{tabularx}
\label{P1:simresults}
\end{table}




\begin{table}
\centering
\caption[Marginal and age category specific disparity and residual disparity estimates for best fitting mediator and outcome models.]{Marginal and age category specific disparity and residual disparity estimates for best fitting mediator and outcome models. Disparities estimates are from a frequentist AFT model fit, adjusting for covariates and allowing for a race-age interaction. Residual Disparity measures are estimated from a LDDP, linear, and baseline category logit meditator model. The best fitting model for BMI included a race-sex interaction and the best fitting survival model included race-age interactions. \vspace*{2mm}}
{\small
\begin{tabular}{ccccc}
  \hline\hline
\textbf{Estimate (95\% CI)} & marginal &age $<50$ & $50 \leq$ age $<65$ &  $65\leq$ age\\ 
  \hline\hline
Disparity  &  0.83 (0.71, 0.98) & 0.64 (0.47, 0.87) & 1.00  (0.81, 1.46) &0.83 (0.64, 1.09)\\
RD Density &  0.86 (0.73, 1.03)& 0.61 (0.45,0.84)& 1.02 (0.79,1.37)& 0.84 (0.67,1.09)\\
RD Linear &  0.82 (0.70, 0.97)& 
 0.61 (0.45, 0.83) & 1.08 (0.77. 1.38)& 0.84 (0.65, 1.10)\\
RD BCL     & 0.85 (0.72, 1.03) & 0.61 (0.46, 0.83) & 1.00 (0.76, 1.33) & 0.86 (0.67, 1.13) \\ 
   \hline
\end{tabular}}
\label{P1:tableBEST}
\end{table}


\end{document}